\documentclass[11pt,a4paper]{article}
\usepackage{jcappub}
\usepackage{bm}
\usepackage{booktabs}
\usepackage{dcolumn}
\usepackage{array}
\usepackage{comment}
\usepackage{placeins}
\usepackage{xcolor}

\title{Strong-lensing effects in high-redshift massive black-hole binary population inference}

\author[a,b,c]{Yan-Heng Jin}
\author[d,a,e]{Wen-Biao Han}

\affiliation[a]{School of Fundamental Physics and Mathematical Sciences, Hangzhou Institute for Advanced Study,
University of Chinese Academy of Sciences, Hangzhou 310024, China}
\affiliation[b]{Institute of Theoretical Physics, Chinese Academy of Sciences, Beijing 100190, China}
\affiliation[c]{University of Chinese Academy of Sciences, Beijing 100049, China}
\affiliation[d]{State Key Laboratory of Radio Astronomy and Technology, Shanghai Astronomical Observatory,
Chinese Academy of Sciences, 80 Nandan Road, Shanghai 200030, China}
\affiliation[e]{School of Astronomy and Space Science, University of Chinese Academy of Sciences, Beijing 100049, China}

\emailAdd{jinyanheng25@mails.ucas.ac.cn}
\emailAdd{wbhan@shao.ac.cn}

\abstract{
High-redshift massive black-hole binary (MBHB) mergers provide a probe of black-hole seed formation and early galaxy assembly, but the population detected by LISA can be modified by galaxy-scale strong lensing. We quantify this effect for MBHB mergers at $10\leq z\leq20$ and assess its impact on the inference of intrinsic formation-channel fractions. We use four channels, corresponding to light- and heavy-seed scenarios with delayed and non-delayed mergers, and compute 4-year LISA-detectable event numbers with and without strong lensing. To bracket the uncertain lens population, we compare a conservative velocity-dispersion-function (VDF) prescription with an optimistic halo-mass-function (HMF)-based prescription. We consider a reference mixture with equal seed-channel weights and additional mixtures in which the intrinsic seed population is weighted toward selected formation channels, and jointly infer the formation fractions and the strength of the lensing contribution. Strong lensing does not affect all channels equally: it can change the detected channel mixture as well as the total number of detections. This effect is weak for the conservative VDF prescription, but becomes significant in the high-lensing-rate HMF case, where neglecting strong lensing can bias the recovered formation fractions. The inference precision depends on the underlying intrinsic channel composition, while the lensing contribution is more accurately recovered in the high-lensing-rate case. These results indicate that galaxy-scale strong-lensing effects and event-count information should be included when using high-redshift MBHB detections to infer intrinsic formation-channel fractions.
}


\begin{document}
\maketitle

	\section{Introduction}
	
	Massive black holes are observed in the centers of a large fraction of galaxies, yet their origin and early growth remain uncertain. Two broad classes of seed models are commonly considered: light seeds, formed as remnants of Population III stars~\cite{PhysRevD.93.024003,Regan_2024,Liu_2024}, and heavy seeds~\cite{PhysRevD.93.024003,Regan_2024,Wise_2019}, formed through the collapse of protogalactic gas or related dynamical processes. These scenarios predict different merger histories, mass distributions, redshift distributions, and delay times between galaxy mergers and massive black-hole binary (MBHB) coalescences. Gravitational-wave observations of high-redshift MBHBs therefore provide a direct probe of black-hole seed formation and early galaxy assembly~\cite{Sesana_2011,Katz_2019}.
	
	In this work we use four representative MBHB populations based on the semianalytic models summarized in Ref.~\cite{Barausse_2012,Barausse_2023}. These models include light-seed and heavy-seed scenarios as well as different assumptions about the delays between galaxy mergers and MBHB coalescences. We use B12 populations as representative no-delay channels and K16 populations as representative delayed channels, giving four formation channels: LS-d, LS-nod, HS-d, and HS-nod. This setup allows us to ask whether future gravitational-wave observations can recover the relative contribution of seed type and delay physics in the high-redshift Universe~\cite{Barkana_2000,davari2026gravitationalwavescosmicdawn,Sereno_2011}.
	
	The detected MBHB population is primarily affected by detector selection~\cite{Mandel_2019}. A detector observes only signals whose signal-to-noise ratio exceeds a chosen threshold, so the observed sample depends on source mass, redshift, spin, sky position, and orientation. Gravitational lensing can further modify this population~\cite{Chen_2026,Liao_2017,Bartelmann_2010,Dai_2017}. While weak lensing~\cite{Oguri2018} mainly perturbs inferred distances and source-frame masses, and microlensing~\cite{_al_kan_2023,Takahashi_2003,Yuan_2026} can introduce waveform-level distortions, we focus here on strong lensing by galaxy-scale lenses. Strong lensing~\cite{Sereno_2010,Robertson_2020,Ezquiaga_2021,Dai_2017,Oguri2018} can magnify otherwise subthreshold sources and generate multiple detectable images of the same intrinsic merger, thereby changing the detected high-redshift MBHB sample.
	
	The strong-lensing contribution depends sensitively on the population of lenses at intermediate and high redshift, which is poorly constrained~\cite{Yue_2022}. To bracket this uncertainty, we consider two lens-population prescriptions. The first follows the velocity-dispersion-function approach of Oguri~\cite{Oguri2018}, in which a locally measured galaxy velocity-dispersion function is evolved with redshift using simulation-based information; we use it as a conservative low-lensing prescription. The second is based on a halo-mass-function construction, where dark-matter halo abundances are mapped to an effective velocity dispersion; this gives a higher optical depth and is used as an optimistic high-lensing prescription~\cite{maityStrongLensingCosmography2026}. We do not regard the HMF-based prescription as a calibrated galaxy-lens model, but rather as a way to test the sensitivity of the inference to lens-population uncertainty.
	
	Galaxy-scale strong lensing can affect MBHB population inference because magnification may change the detectable contribution of different formation channels unequally. Whether this correction is negligible or important depends on the lensing rate and on how each formation channel is distributed in source parameters and signal-to-noise ratio. A weak lensing contribution may only slightly perturb the detected population, whereas a stronger contribution can alter the detected channel mixture and bias the recovered intrinsic fractions if it is ignored. We therefore quantify the effect under both conservative and optimistic lens-population prescriptions.
	
	We test this effect in three steps. First, we compute 4-year LISA-detectable event numbers with and without strong lensing for the four MBHB formation channels over $10 \leq z \leq 20$. LISA~\cite{AmaroSeoane2017LISA,Colpi2024LISA} is used as the reference detector for space-based high-redshift MBHB observations; since Taiji~\cite{Hu:2017mde,Luo:2021qji,Ruan_2020,Liu_2023} has a similar mHz sensitivity band, the LISA results are expected to be qualitatively representative of Taiji. Second, we infer the intrinsic formation-channel fractions, first without lensing and then with strong lensing included in the mock observed catalogues and inference models. This comparison separates the recoverability of the formation fractions from the additional bias or precision change introduced by lensing. Third, we extend the inference to include an effective lensing-rate parameter and test whether the data can jointly constrain the formation fractions and the strong-lensing contribution. We perform this analysis for a reference mixture with equal seed-channel weights and for additional mixtures in which the intrinsic seed population is weighted toward selected formation channels.
	
	The paper is organized as follows. Section~\ref{sec:framework} describes the source populations, lens populations, detector selection, and inference model. Section~\ref{sec:event-numbers} presents the event-number calculation. Section~\ref{sec:population-inference} presents the inference of formation fractions and the joint inference of formation fractions and the lensing-rate parameter. Section~\ref{sec:discussion-conclusions} gives the discussion and conclusions.
	
	\section{Framework}
	\label{sec:framework}
	
	This section defines the forward model used throughout the paper. We first specify the MBHB source populations and the formation-channel parameters. We then describe the lens-population prescriptions used to compute galaxy-scale strong-lensing optical depths, the detector selection used to construct detectable samples, and the Bayesian model used to infer intrinsic channel fractions from the detected-event parameter distributions.

	\subsection{Source populations}
	\label{subsec:source-populations}
	
	We construct the intrinsic MBHB population from the semianalytic catalogues of Refs.~\cite{Barausse_2012,Barausse_2023}. Each catalogue is interpreted as a single-channel population, namely the merger history obtained if the simulated Universe followed that formation scenario alone. We combine these catalogues into mixed populations in which multiple channels contribute simultaneously. We consider four representative channels,
	\begin{equation}
		{\rm LS\mbox{-}d},\quad {\rm LS\mbox{-}nod},\quad
		{\rm HS\mbox{-}d},\quad {\rm HS\mbox{-}nod}.
	\end{equation}
	Here LS and HS denote light- and heavy-seed scenarios, while d and nod denote delayed and non-delayed MBHB mergers. We use the K16 catalogues as representative delayed channels and the B12 catalogues as representative non-delayed channels. Specifically, LS-d is represented by the popIII-d K16 catalogue, HS-d by the Q3-d K16 catalogue, and the non-delayed light- and heavy-seed channels by the corresponding B12 catalogues.
	
\begin{figure}[t]
		\centering
		\includegraphics[width=0.7\linewidth]{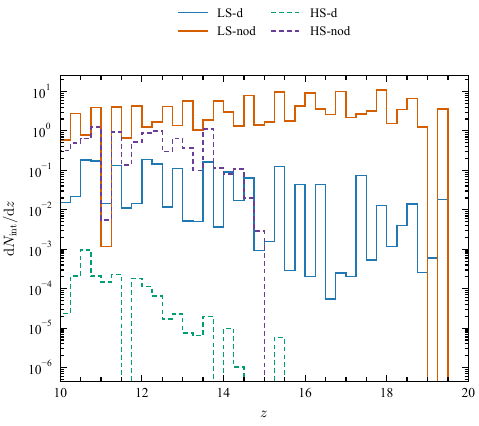}
		\caption{\label{fig:source-dndz}
			Intrinsic redshift distributions of the four MBHB source populations at $10\leq z\leq20$. The curves show the weighted $dN/dz$ distributions of the four formation-channel catalogues for a 4-year observing time, before detector selection and lensing effects are applied.}
	\end{figure}

	For the event-number and inference analyses, we restrict the source population to $10\leq z\leq20$ and normalize all intrinsic and detectable event numbers to a 4-year observing time. Figure~\ref{fig:source-dndz} shows the intrinsic redshift distributions of the four single-channel catalogues before detector and lensing effects are applied.
    
    Let $\Lambda^{\rm cat}_{{\rm int},k}$ denote the weighted intrinsic number of mergers in catalogue $k$ over this redshift interval and observing time. This quantity reflects the model-dependent formation and merger yield of channel $k$, rather than its relative abundance in the real Universe. Its numerical values are reported together with the corresponding LISA-detectable event numbers in Table~\ref{tab:single-channel-event-numbers}.
	
	We introduce normalized seed-channel weights $a_k$, with $\sum_k a_k=1$, to describe how the intrinsic seed population is assigned among the four formation channels before their different merger yields are taken into account. The mixed-population intrinsic count is
	\begin{equation}
		\Lambda_{{\rm int},k}
		=
		a_k \Lambda^{\rm cat}_{{\rm int},k}.
		\label{eq:mixed-intrinsic-count}
	\end{equation}
	This construction uses the weighted nature of the catalogues: event numbers are obtained by summing event weights, and $a_k$ acts as an overall rescaling of the weights in catalogue $k$. It changes the total contribution of that channel while leaving its normalized redshift, mass, spin, and other source-parameter distributions unchanged. Thus the mixture is a linear combination of precomputed single-channel catalogues, not a new self-consistent population-synthesis calculation.
	
	The intrinsic channel fractions are then
	\begin{equation}
		F_k =
		\frac{a_k\Lambda^{\rm cat}_{{\rm int},k}}
		{\sum_j a_j\Lambda^{\rm cat}_{{\rm int},j}} ,
		\label{eq:intrinsic-fraction}
	\end{equation}
	where $k$ runs over the four channels. Thus the final intrinsic merger fraction is determined by both the seed-channel weight and the single-channel merger yield. For fixed catalogue yields $\Lambda^{\rm cat}_{{\rm int},k}$, the weights $a_k$ and the intrinsic fractions $F_k$ are equivalent ways of specifying the mixed population. We further reparameterize the intrinsic fractions as~\cite{Shen_2026}
	\begin{align}
		f_1 &= F_{\rm LS\mbox{-}d}+F_{\rm LS\mbox{-}nod}, \\
		f_2 &= \frac{F_{\rm LS\mbox{-}d}}
		{F_{\rm LS\mbox{-}d}+F_{\rm LS\mbox{-}nod}}, \\
		f_3 &= \frac{F_{\rm HS\mbox{-}d}}
		{F_{\rm HS\mbox{-}d}+F_{\rm HS\mbox{-}nod}} .
	\end{align}
	Here $f_1$ is the total light-seed fraction, while $f_2$ and $f_3$ describe the delayed fraction within the light- and heavy-seed sectors.
	
	The mixture definitions used in this work are listed in Table~\ref{tab:mixture-definitions}. The reference mixture assigns equal seed-channel weights to the four channels. We also consider mixtures weighted toward selected channels to test whether the conclusions drawn from the reference case remain valid when the intrinsic seed population is shifted toward different formation pathways. For each weighted family, the mild and intermediate cases set the selected channel weight to be 10 and 100 times larger than the other three weights, respectively. The extreme case assigns a sufficiently large weight to the selected channel that its intrinsic merger fraction becomes $F_k\simeq0.9$.
	
	\begin{table}[t]
	\centering
	\footnotesize
		\caption{\label{tab:mixture-definitions}
			Intrinsic mixture definitions used in this work. The vector $\bm{a}$ is ordered as
			$(a_{\rm LS\mbox{-}d},a_{\rm LS\mbox{-}nod},a_{\rm HS\mbox{-}d},a_{\rm HS\mbox{-}nod})$.
			The corresponding intrinsic-fraction vector is $(f_1,f_2,f_3)$. In each weighted family, mild and intermediate denote selected-channel weights 10 and 100 times larger than the other weights, while extreme denotes a case with $F_k\simeq0.9$ for the selected channel.}
		
			\begin{tabular*}{\textwidth}{@{\extracolsep{\fill}}lll@{}}
				Mixture & $\bm{a}$ & $(f_1,f_2,f_3)$ \\
				\hline
				Reference
				& $(0.25,0.25,0.25,0.25)$
				& $(0.94184,0.01205,2.39\times10^{-4})$ \\
				LS-d weighted, mild
				& $(0.76923,0.07692,0.07692,0.07692)$
				& $(0.94723,0.10871,2.39\times10^{-4})$ \\
				LS-d weighted, intermediate
				& $(0.97087,0.00971,0.00971,0.00971)$
				& $(0.97261,0.54944,2.39\times10^{-4})$ \\
				LS-d weighted, extreme
				& $(0.99619,0.00127,0.00127,0.00127)$
				& $(0.99412,0.90536,2.39\times10^{-4})$ \\
				HS-nod weighted, mild
				& $(0.07692,0.07692,0.07692,0.76923)$
				& $(0.61827,0.01205,2.39\times10^{-5})$ \\
				HS-nod weighted, intermediate
				& $(0.00971,0.00971,0.00971,0.97087)$
				& $(0.13941,0.01204,2.39\times10^{-6})$ \\
				HS-nod weighted, extreme
				& $(0.00672,0.00672,0.00672,0.97984)$
				& $(0.09998,0.01205,1.64\times10^{-6})$ \\
			\end{tabular*}
		
	\end{table}
	
	We do not attempt an exhaustive scan of all possible seed-channel weights. Mixtures further weighted toward LS-nod are not shown separately because LS-nod already has by far the largest single-channel merger yield and dominates the intrinsic reference mixture. Further increasing its weight would therefore produce behavior close to the reference case. Mixtures weighted toward HS-d are also not shown because the HS-d catalogue yield is extremely small: moderate increases in its weight do not appreciably change the mixed population, whereas an extreme weighting leaves too few detectable events for an informative inference test.
	
	For each channel, the catalogue provides source-frame binary parameters and event weights, from which we construct the detectable samples and the detected-event probability densities used in the population inference. In the LISA inference we retain all four channels and infer $(f_1,f_2,f_3)$.
	
	\subsection{Lens populations}
	\label{subsec:lens-populations}
	
	The strong-lensing optical depth is determined by the mass model of individual lenses and by the population model of the deflectors. The former specifies the density profile and geometry of a single lens, while the latter gives the abundance of lenses as a function of redshift and velocity dispersion. In this work, we model individual lenses as singular isothermal ellipsoids with external shear (SIE+shear)~\cite{Kormann1994SIE,OguriMarshall2010} and keep this lens mass model fixed throughout. We then use different velocity-dispersion prescriptions to bracket the uncertainty in the lens population.
	
	For a source at redshift $z_s$, the optical depth is computed as
	\begin{equation}
		\tau(z_s)
		=
		\int_0^{z_s} dz_l\,
		\frac{dV_c}{dz_l d\Omega}
		\int d\sigma\,
		\phi(\sigma,z_l)\,
		S_{\rm SIE+\gamma}(\sigma,z_l,z_s).
		\label{eq:optical-depth}
	\end{equation}
	Here $z_l$ is the lens redshift, $dV_c/(dz_l d\Omega)$ is the comoving volume element per unit redshift and solid angle, $\sigma$ is the lens velocity dispersion, $\phi(\sigma,z_l)\equiv d^2N/(d\sigma dV_c)$ is the comoving VDF, and $S_{\rm SIE+\gamma}$ is the SIE+shear strong-lensing cross section expressed as an angular area in the source plane. For isothermal lenses the cross section scales approximately as $\sigma^4$, so the high-$\sigma$ tail of the VDF can dominate the optical depth.
	
	\paragraph*{Oguri VDF.}
	
	Our conservative lens-population prescription follows the galaxy VDF model used by Oguri~\cite{Oguri2018}. The local VDF~\cite{Bernardi_2010} is
	\begin{equation}
		\phi_{\rm loc}(\sigma)
		=
		\phi_\ast
		\left(\frac{\sigma}{\sigma_\ast}\right)^\alpha
		\exp\left[-\left(\frac{\sigma}{\sigma_\ast}\right)^\beta\right]
		\frac{\beta}{\Gamma(\alpha/\beta)}
		\frac{1}{\sigma},
		\label{eq:oguri-local-vdf}
	\end{equation}
	where $\phi_\ast = 2.099 \times 10^{-2} (h/0.7)^3\,\mathrm{Mpc}^{-3}$ is the normalization, $\sigma_\ast = 113.78\,\mathrm{km\,s^{-1}}$ is the characteristic velocity dispersion, $\alpha=0.94$ and $\beta=1.85$ control the low-$\sigma$ and high-$\sigma$ behavior, and $\Gamma$ is the gamma function. Redshift evolution is included through
	\begin{equation}
		\phi_{\rm Oguri}(\sigma,z_l)
		=
		\phi_{\rm loc}(\sigma)
		\frac{\phi_{\rm hyd}(\sigma,z_l)}
		{\phi_{\rm hyd}(\sigma,0)} .
		\label{eq:oguri-evolving-vdf}
	\end{equation}
	Here $\phi_{\rm hyd}(\sigma,z_l)$~\cite{Torrey_2015} is the VDF measured from the hydrodynamical simulation at redshift $z_l$, and $\phi_{\rm hyd}(\sigma,0)$ is the corresponding simulated VDF at $z_l=0$. Their ratio gives a dimensionless redshift-evolution factor, while the normalization at $z_l=0$ remains fixed by the observed local galaxy VDF. We therefore use this model as the conservative, low-lensing-rate prescription. In our calculation we restrict the velocity dispersion to $30\,{\rm km\,s^{-1}}\leq \sigma \leq 600\,{\rm km\,s^{-1}}$. Within this range, the adopted Oguri VDF prescription does not provide lens-population support beyond $z_l\simeq14$. This behavior follows from the evolving VDF prescription itself, rather than from an additional redshift cutoff imposed in the lensing calculation.
	
	\paragraph*{HMF-derived effective VDF.}
	
	As an optimistic alternative, we construct an effective VDF from the halo mass function (HMF), following halo-based strong-lensing forecasts~\cite{maityStrongLensingCosmography2026}. Starting from the comoving halo abundance $d^2N/(dM_h dV_c)$, we map halo mass to velocity dispersion with the virial relations
	\begin{equation}
		M_h=\frac{4\pi}{3}R_h^3\rho_h(z_l),
		\qquad
		\sigma^2=\frac{G M_h}{R_h},
		\label{eq:hmf-virial}
	\end{equation}
	where $M_h$ is the halo mass, $R_h$ is the virial radius, $\rho_h(z_l)$ is the characteristic virial density, and $G$ is Newton's constant. The effective VDF is then
	\begin{equation}
		\phi_{\rm HMF}(\sigma,z_l)
		\equiv
		\frac{d^2N}{d\sigma dV_c}
		=
		\frac{d^2N}{dM_h dV_c}
		\frac{dM_h}{d\sigma},
		\label{eq:hmf-vdf}
	\end{equation}
	with
	\begin{equation}
		\frac{dM_h}{d\sigma}=\frac{3M_h}{\sigma}.
		\label{eq:hmf-jacobian}
	\end{equation}
	This $\phi_{\rm HMF}$ should be interpreted as a halo-based effective VDF rather than a directly observed galaxy VDF. Since it includes a broader population of possible high-redshift deflectors, it generally predicts a larger optical depth and is used here as the optimistic, high-lensing-rate prescription.
	
\begin{figure}[tbp]
		\centering
		\includegraphics[width=1\linewidth]{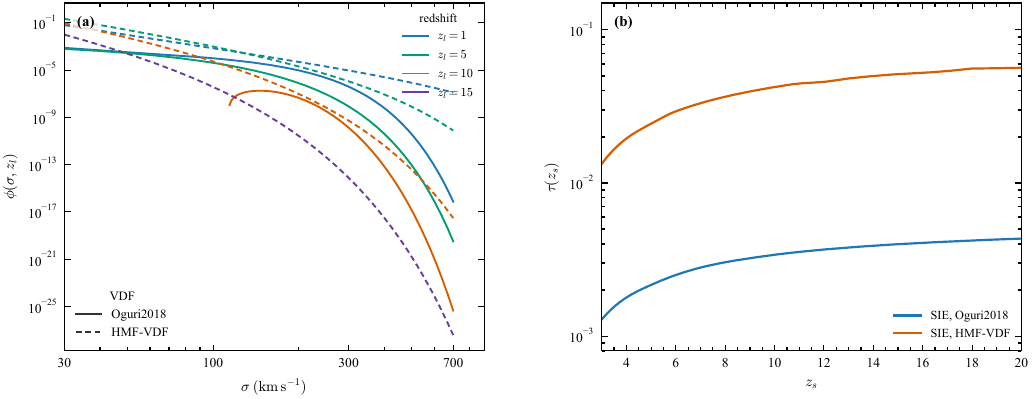}
		\caption{\label{fig:lens-populations}
			Lens-population prescriptions used in this work. The left panel compares $\phi(\sigma,z_l)$ for the conservative Oguri VDF prescription (solid lines) and the optimistic HMF-derived effective VDF prescription (dashed lines) at representative lens redshifts. The absence of the Oguri curve at $z_l=15$ reflects the $z_l\simeq14$ upper-redshift limit of the adopted Oguri VDF prescription over the velocity-dispersion range used here. The right panel shows the corresponding SIE+shear strong-lensing optical depth $\tau(z_s)$ as a function of source redshift for the two VDF prescriptions.}
	\end{figure}

	In both prescriptions, the optical depth is computed with the same SIE+shear lensing cross section. Differences between the Oguri and HMF-based results therefore reflect the assumed abundance of lenses under a fixed lens mass model. Figure~\ref{fig:lens-populations} compares the two VDF prescriptions and the corresponding SIE+shear optical depths used in this work.
	
	\subsection{Detector selection and lensing effects}
	\label{subsec:detector-selection}
	
\begin{figure}[tbp]
		\centering
		\includegraphics[width=0.7\linewidth]{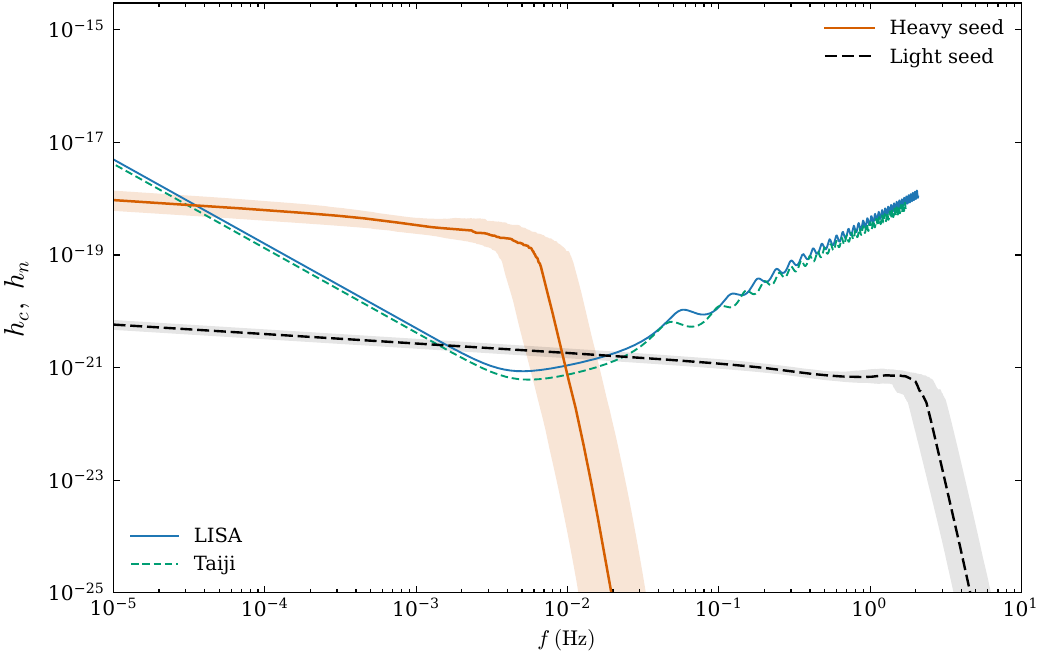}
		\caption{\label{fig:detector-signal-comparison}
			Characteristic noise strains of LISA and Taiji, compared with representative light- and heavy-seed MBHB signals. Detector curves show $h_n(f)=\sqrt{fS_n(f)}$, and source curves show $h_c(f)=2f|\tilde h(f)|$ for nonspinning, face-on IMRPhenomD waveforms. Shaded bands and lines denote the 16th--84th percentile ranges and medians from 250 weight-sampled binaries in each seed family.}
	\end{figure}

	We next define how the intrinsic MBHB catalogues are converted into detectable samples. For each merger event, we first compute the signal-to-noise ratio it would have in the absence of lensing, denoted by $\rho_0$. This quantity depends on the source parameters, redshift, sky location, orientation, waveform model, detector response, and detector sensitivity. For LISA, we use the time-dependent detector response~\cite{Guti_rrez_2025}, which accounts for the orbital motion of the constellation and the corresponding modulation of the observed waveform. Figure~\ref{fig:detector-signal-comparison} illustrates the close mHz-band sensitivities of LISA and Taiji, motivating our use of LISA as the reference space-based detector.
	
	Starting from the intrinsic catalogue of each formation channel, we generate the corresponding lensed catalogue by solving the SIE+shear lens equation. For each bin in source redshift, we sample deflectors from the adopted VDF prescription and assign ellipticities and external shears to them~\cite{Oguri_2010,Oguri_2021}. We then solve the lens equation to obtain the image multiplicities, magnifications, and time delays. Each lensed image corresponds to a waveform arriving at the detector with its own magnification and arrival time; in the event-number calculation, each detectable image is counted as an observed gravitational-wave event. The resulting events are then filtered by the signal-to-noise-ratio threshold to obtain the events that enter the detectable sample.
	
	We adopt the detection threshold
	\begin{equation}
		\rho_{\rm obs} \geq \rho_{\rm th},
		\qquad
		\rho_{\rm th}=8 .
		\label{eq:snr-threshold}
	\end{equation}
	For an unlensed event, $\rho_{\rm obs}=\rho_0$. For a lensed image with magnification $\mu$~\cite{Dai_2017}, the gravitational-wave strain amplitude is enhanced by $\sqrt{\mu}$, and therefore
	\begin{equation}
		\rho_{\rm obs}
		=
		\begin{cases}
			\rho_0, & \text{unlensed event},\\
			\sqrt{\mu}\rho_0, & \text{lensed image}.
		\end{cases}
		\label{eq:observed-snr}
	\end{equation}
	Thus, a lensed image is detectable if $\sqrt{\mu}\rho_0\geq\rho_{\rm th}$. Strong lensing can therefore move intrinsically faint or high-redshift sources from below the threshold into the detectable sample.
	
	After applying these criteria to the propagated catalogues, we obtain the expected number of detectable events in each formation channel, denoted by $\Lambda_{{\rm det},k}$. All $\Lambda_{{\rm det},k}$ values are normalized to the same 4-year observing time as the intrinsic counts. The corresponding detection efficiency is
	\begin{equation}
		\epsilon_k
		=
		\frac{\Lambda_{{\rm det},k}}
		{\Lambda_{{\rm int},k}},
		\label{eq:detection-efficiency}
	\end{equation}
	where $\Lambda_{{\rm int},k}$ is the weighted intrinsic number of mergers in channel $k$ over the same redshift interval and observing time. The efficiency $\epsilon_k$ measures the fraction of intrinsic mergers in channel $k$ that enter the detectable sample after lensing propagation and detector selection.
	
	For each channel, we also construct the normalized parameter distribution of detectable events,
	\begin{equation}
		P_k(\theta)
		=
		\frac{1}{\Lambda_{{\rm det},k}}
		\frac{d\Lambda_{{\rm det},k}}{d\theta},
		\qquad
		\int d\theta\,P_k(\theta)=1 .
		\label{eq:detected-template}
	\end{equation}
	Here $\theta$ denotes the event parameters used in the population inference, such as redshift, chirp mass, mass ratio, and effective spin. The distribution $P_k(\theta)$ is constructed from the detectable catalogue of channel $k$ and represents the probability density for a detected event from that channel to have parameters $\theta$.
	
\begin{figure}[tbp]
		\centering
		\includegraphics[width=\linewidth]{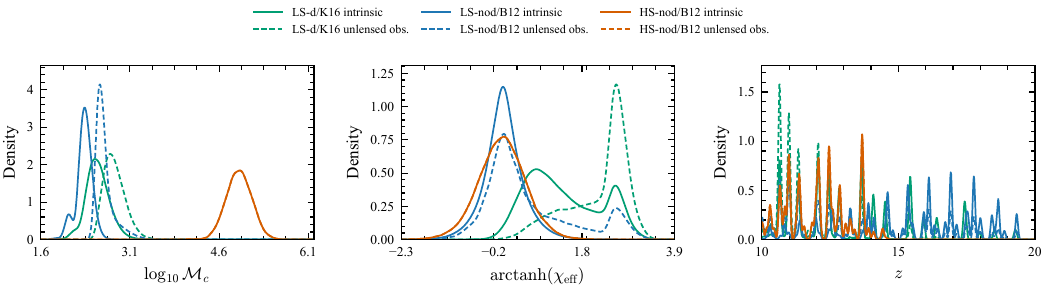}
		\vspace{0.3em}
		\includegraphics[width=\linewidth]{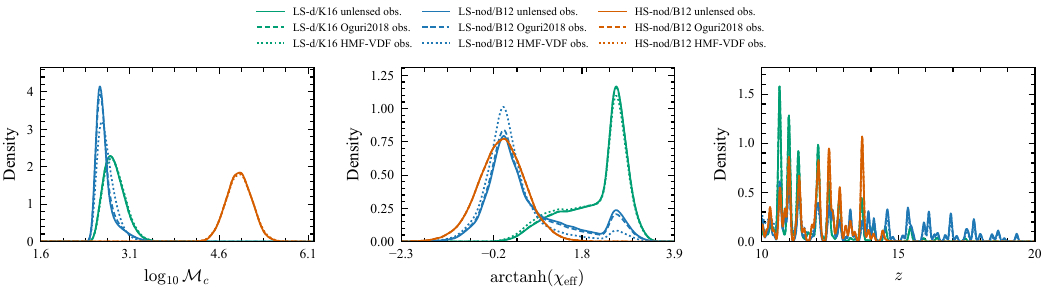}
		\caption{\label{fig:population-distributions}
			Population distributions for the MBHB formation channels in the reference mixture. The upper row compares the intrinsic distributions with the unlensed LISA-detectable distributions. The lower row compares the unlensed detectable distributions with the lensed distributions obtained with the Oguri-VDF and HMF-VDF prescriptions.}
	\end{figure}

	Figure~\ref{fig:population-distributions} illustrates how the intrinsic channel distributions are reshaped by LISA detectability and by strong lensing. The upper panels compare the intrinsic source distributions with the unlensed detectable distributions, while the lower panels compare the unlensed detectable distributions with the lensed detectable distributions. These differences show that the population inference must be built from detected-event distributions rather than from intrinsic source distributions alone.
	
	\subsection{Inference model}
	\label{subsec:inference-model}
	
	The mock observed catalogues are chosen according to the analysis being performed. In the inference without lensing we use the unlensed LISA-detectable catalogues as mock observed data, while in the lensing analysis we use the lensed detectable catalogues. The inference model can then be built from either unlensed or lensed channel distributions, allowing us to test whether ignoring strong lensing biases the recovery of the intrinsic formation-channel fractions. The event parameters used in the inference are~\cite{Breivik_2016}
	\begin{equation}
		\theta =
		\left(
		\log_{10}{\cal M}_c,\,
		z,\,
		\operatorname{arctanh}\chi_{\rm eff}
		\right),
		\label{eq:inference-space}
	\end{equation}
	where ${\cal M}_c$ is the chirp mass and $\chi_{\rm eff}$ is the effective spin.
	
	We assume that the relevant formation channels in $10\leq z\leq20$ are known. For each channel $k$ and model choice $X$, we also assume that the detected-event probability density $P^X_k(\theta)$ and the detection efficiency $\epsilon^X_k$ are specified. Here $X={\rm unlen}$ denotes distributions constructed without strong lensing, while $X={\rm len}$ denotes distributions constructed from the lensed catalogues. In real observations, the channel set, the distributions $P^X_k(\theta)$, and the efficiencies $\epsilon^X_k$ would have to be inferred or compared within a broader hierarchical model~\cite{Mandel_2019,langen2025hierarchicalbayesianinferenceanalytical,Thrane_2019}.
	
	Let $p^X_{kb}$ be the binned probability for a detected event from channel $k$ to fall in bin $b$. For intrinsic channel fractions $F_k(\bm{f})$, the detected channel fraction predicted by model $X$ is
	\begin{equation}
		\pi^X_k(\bm{f})
		=
		\frac{F_k(\bm{f})\epsilon^X_k}
		{\sum_j F_j(\bm{f})\epsilon^X_j},
		\label{eq:detected-channel-fraction}
	\end{equation}
	and the predicted bin probability is
	\begin{equation}
		q^X_b(\bm{f})
		=
		\sum_k \pi^X_k(\bm{f}) p^X_{kb}.
		\label{eq:mixture-template}
	\end{equation}
	
	The conditional likelihood uses only the shape of the observed distribution,
	\begin{equation}
		\ln {\cal L}_{\rm cond}
		=
		\sum_b n_b \ln q^X_b(\bm{f}),
		\label{eq:conditional-likelihood}
	\end{equation}
	where $n_b$ is the bin count of the corresponding mock observed data.
	
	We also use a Poisson likelihood that includes the total event count. We define the total intrinsic number of MBHB mergers as
	\begin{equation}
		A=\sum_k \Lambda_{{\rm int},k},
		\label{eq:total-intrinsic-number}
	\end{equation}
	where the intrinsic counts are normalized to a 4-year observing time. The expected count in bin $b$ is
	\begin{equation}
		\lambda^X_b(\bm{f},A)
		=
		\sum_k
		\frac{A F_k(\bm{f})}{\Lambda_{{\rm int},k}}
		H^X_{kb},
		\label{eq:poisson-expected-count}
	\end{equation}
	where $H^X_{kb}$ is the unnormalized weighted number of detectable events from channel $k$ in bin $b$ under model $X$, with $\sum_b H^X_{kb}=\Lambda^X_{{\rm det},k}$. The Poisson log likelihood is
	\begin{equation}
		\ln {\cal L}_{\rm Pois}
		=
		\sum_b
		\left[
		n_b\ln \lambda^X_b
		-
		\lambda^X_b
		\right],
		\label{eq:poisson-likelihood}
	\end{equation}
	with the constant term $-\sum_b\ln(n_b!)$ omitted.
	
	We consider known-$A$ and sampled-$A$ cases. The known-$A$ case fixes $A$ to the input intrinsic merger number and gives the theoretical precision attainable when this number is known. The sampled-$A$ case treats $A$ as an additional free parameter and tests whether the Poisson event-count information still improves the recovery of the channel fractions when the overall intrinsic normalization is uncertain. In the main results below, we focus on LISA and infer $(f_1,f_2,f_3)$. The weighted mixtures are presented for the known-$A$ Poisson likelihood with the lensed inference model, which allows us to examine how the inference depends on the intrinsic seed-channel weighting in the best-constrained setup. The lensing-rate analysis is treated by extending the parameter set to include a lensing-rate parameter $R$.

	\section{Event numbers}
	\label{sec:event-numbers}
	
	We first examine how LISA detectability and galaxy-scale strong lensing change the expected number of MBHB events in each single-channel catalogue. These single-channel results are the event-number building blocks for the reference and weighted mixtures defined in Sec.~\ref{subsec:source-populations}. In the lensed cases, each detectable lensed image is counted as an observed gravitational-wave event. All event numbers are weighted expected counts and need not be integers. The corresponding detectable event numbers for the mixed populations are collected in Appendix~\ref{app:inference-results}.
	
	\begin{table}[t]
	\centering
	\setlength{\tabcolsep}{10pt}
	\caption{\label{tab:single-channel-event-numbers}
    Single-channel intrinsic merger yields and LISA-detectable event
    numbers for $10\leq z\leq20$ over 4 years. The detectable columns
    give the unlensed, Oguri-VDF, and HMF-VDF cases.}
		
			\begin{tabular}{lcccc}
				Channel
				& $\Lambda^{\rm cat}_{{\rm int},k}$
				& unlen
				& Oguri
				& HMF \\
				\hline
				LS-d   & 682.20   & 51.18   & 51.88   & 60.15 \\
				LS-nod & 55934.55 & 97.00   & 112.66  & 299.24 \\
				HS-d   & 0.84     & 0.83    & 0.84    & 0.87 \\
				HS-nod & 3495.38  & 3493.68 & 3507.62 & 3669.56 \\
			\end{tabular}
		
	\end{table}
	
	Table~\ref{tab:single-channel-event-numbers} shows that LISA detects the four channels with very different efficiencies. The heavy-seed channels are selected with high efficiency: almost all HS-nod events and nearly all HS-d events are above threshold, because their larger masses produce stronger signals in the LISA band. The light-seed channels are less efficiently selected. This is especially important for LS-nod, which has the largest intrinsic yield but a much smaller detectable fraction.
	
	Strong lensing increases the detectable event numbers by magnifying otherwise subthreshold sources. The increase is small for the Oguri VDF and larger for the HMF-derived effective VDF. The largest absolute enhancement occurs in LS-nod, because this channel contains many subthreshold events that can be promoted into the detectable sample. The heavy-seed channels are already efficiently detected without lensing, so lensing produces a smaller relative change.
	
	Thus, strong lensing does not rescale all channels by the same factor. Its impact depends on the source distribution, the number of near-threshold events, and the adopted lens-population prescription. This channel dependence motivates the population-inference analysis below.

	\FloatBarrier

	\section{Population inference}
	\label{sec:population-inference}
	
	We now examine how the event-number behavior translates into population inference. We first infer the formation fractions without lensing, then ask how galaxy-scale strong lensing changes that inference, and finally extend the model to infer the lensing-rate parameter jointly with the formation fractions. The reference-mixture results most directly connected to the main conclusions are presented in this section, while the complete weighted-mixture results are collected in Appendix~\ref{app:inference-results}.

	\subsection{Formation-fraction inference without lensing}
	\label{subsec:f-without-lensing}
	
	We first establish how well the intrinsic formation fractions can be recovered before strong lensing is introduced. In this subsection the mock data are the unlensed LISA-detectable catalogues and the inference uses the corresponding unlensed model. This lensing-free case identifies which parameters are already constrained by the detectable population itself, and which are limited by small channel contributions or by overlap between channel distributions.
	
\begin{figure}[tbp]
		\centering
		\includegraphics[width=0.49\linewidth]{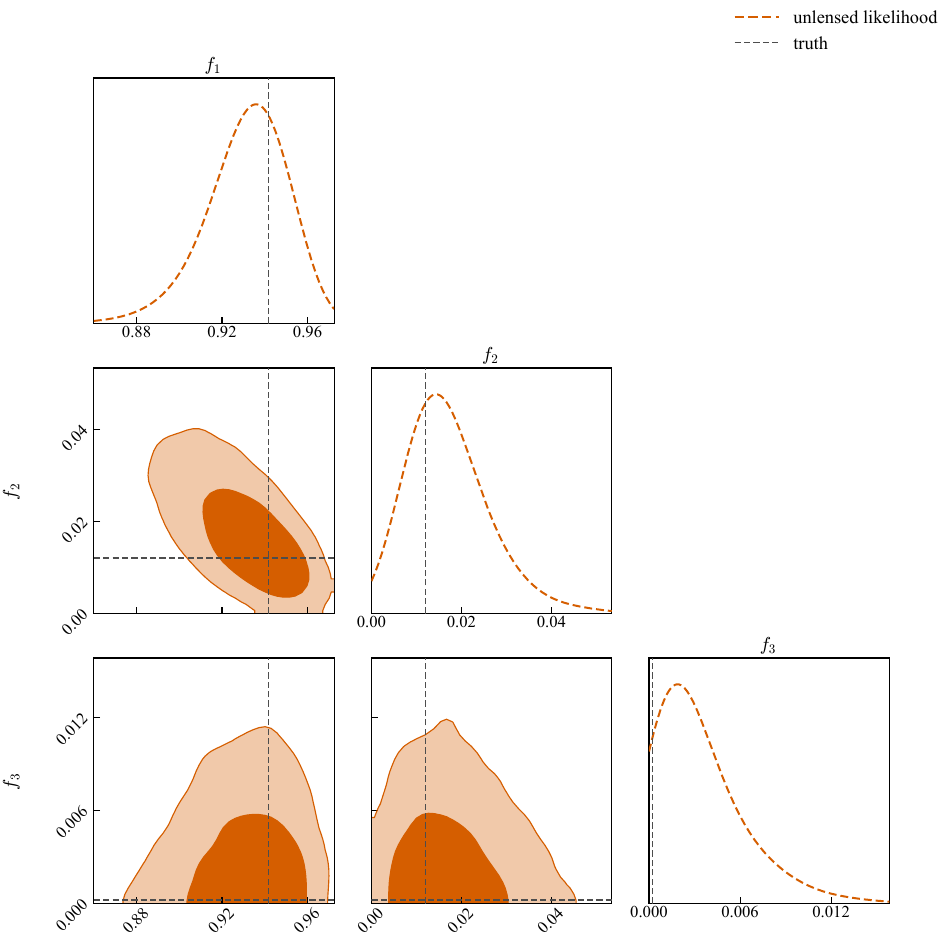}
		\includegraphics[width=0.49\linewidth]{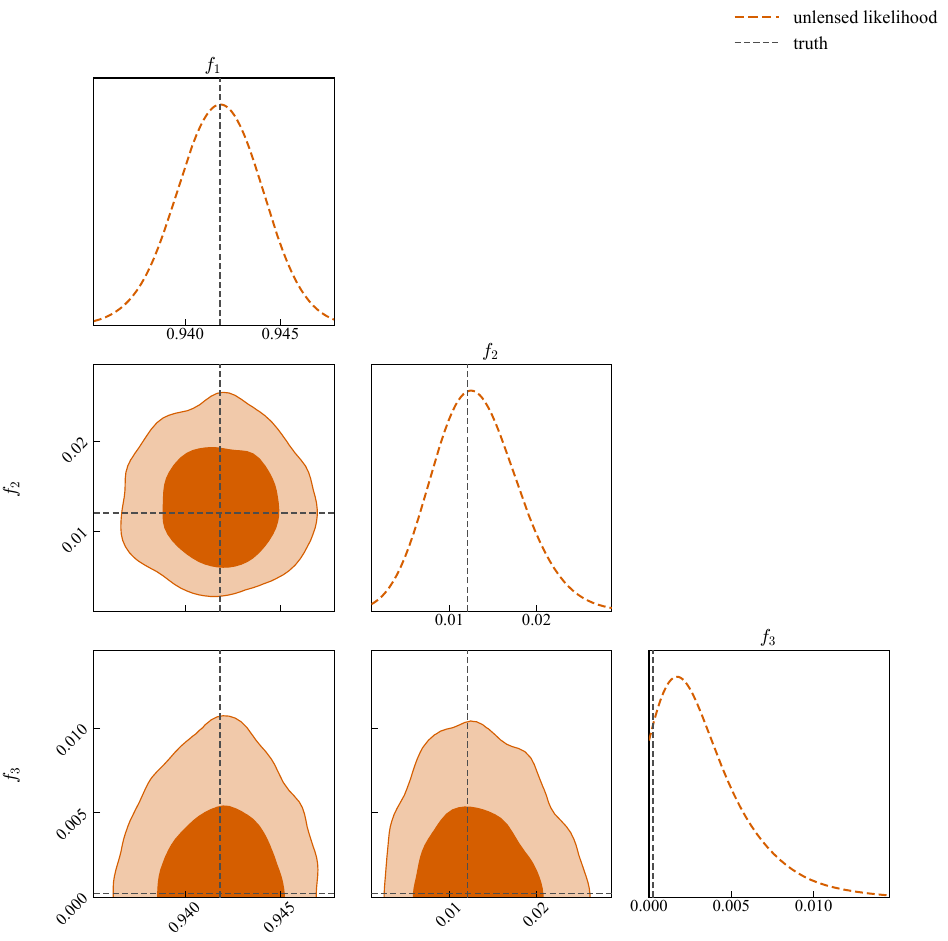}
		\caption{\label{fig:baseline-unlensed}
			Lensing-free LISA formation-fraction inference for the reference mixture. The mock observed data are the unlensed detectable catalogue, and the inference uses the unlensed model. The left panel shows the conditional likelihood, while the right panel shows the known-$A$ Poisson likelihood.}
	\end{figure}

	For the reference mixture, Table~\ref{tab:baseline-reference} compares the three likelihoods defined in Sec.~\ref{subsec:inference-model}, and Fig.~\ref{fig:baseline-unlensed} contrasts the posterior structure for the conditional and known-$A$ cases. The conditional and sampled-$A$ likelihoods give similar constraints, because allowing the total intrinsic event number to vary reduces the additional information carried by the total count. Fixing $A$ gives a much tighter constraint on $f_1$, showing that event-count normalization is informative when it is externally known. The hierarchy among the parameters is also clear: $f_1$ is best recovered, $f_2$ is less precise, and $f_3$ is poorly constrained. This follows from the detected-event distributions in Fig.~\ref{fig:population-distributions}: light- and heavy-seed channels are well separated in chirp mass, whereas delayed and non-delayed channels within the same seed family are harder to distinguish.
	
	The corresponding lensing-free inference for the LS-d- and HS-nod-weighted mixtures is summarized in Table~\ref{tab:baseline-weighted} of Appendix~\ref{app:inference-results}. We show only the known-$A$ likelihood, since the goal is to test how recoverability changes with the underlying seed-channel weighting rather than to repeat the likelihood comparison. The total light-seed fraction $f_1$ remains accurately recovered when the detectable sample contains enough information to separate the light- and heavy-seed sectors. The recovery of $f_2$ depends more strongly on the delayed light-seed contribution: it is recovered within broad but meaningful intervals in the LS-d-weighted mixtures, but becomes much less precise when the light-seed sector is subdominant in the HS-nod-weighted mixtures. The parameter $f_3$ remains difficult in all cases because the HS-d yield is extremely small. Thus, for the population models used here, the heavy-seed sector is effectively HS-nod dominated, and little information is available to resolve a delayed heavy-seed subpopulation.
	
	\begin{table}[tbp]
		\renewcommand{\arraystretch}{1.5}
	\centering
		\caption{\label{tab:baseline-reference}
			Lensing-free LISA inference for the reference mixture. The unlensed detectable catalogue is used as the mock observed data and the inference uses the unlensed model. Entries give posterior medians with 68\% credible intervals.}
		
			\begin{tabular}{lccc}
				Likelihood & $f_1$ & $f_2$ & $f_3$ \\
				\hline
				True value
				& 0.94184
				& 0.01205
				& 0.00024 \\
				conditional
				& $0.93371^{+0.01404}_{-0.01799}$
				& $0.01556^{+0.00905}_{-0.00627}$
				& $0.00238^{+0.00325}_{-0.00170}$ \\
				sampled-$A$
				& $0.93393^{+0.01385}_{-0.01776}$
				& $0.01564^{+0.00834}_{-0.00619}$
				& $0.00209^{+0.00317}_{-0.00152}$ \\
				known-$A$
				& $0.94182^{+0.00189}_{-0.00192}$
				& $0.01278^{+0.00448}_{-0.00384}$
				& $0.00215^{+0.00313}_{-0.00156}$ \\
			\end{tabular}
		
	\end{table}
	
	These lensing-free results set the scale for the lensing analysis below. A poor constraint on a formation fraction is not necessarily caused by strong lensing; it can arise because the relevant channel contributes too few detectable events, or because the channels being compared are not well separated in the inference space. Conversely, parameters that are already well constrained without lensing provide the cleanest tests of whether lensing introduces a bias or improves the precision by adding detectable events.

	\subsection{Formation-fraction inference with lensing}
	\label{subsec:f-with-lensing}
	
	We now examine how galaxy-scale strong lensing changes the formation-fraction inference relative to the lensing-free case above. The preceding analysis shows that the precision of the inference depends on the number of detectable events and on how well the relevant channels can be separated in the inference space. Strong lensing can affect both ingredients by promoting subthreshold sources into the detectable sample and by changing the detected channel mixture. We therefore ask two questions: whether the additional lensed detections improve the recovery of the formation fractions, and whether analyzing lensed data with an unlensed model biases the inferred intrinsic fractions.
	
	\begin{figure}[tbp]
		\centering
		\includegraphics[width=0.49\linewidth]{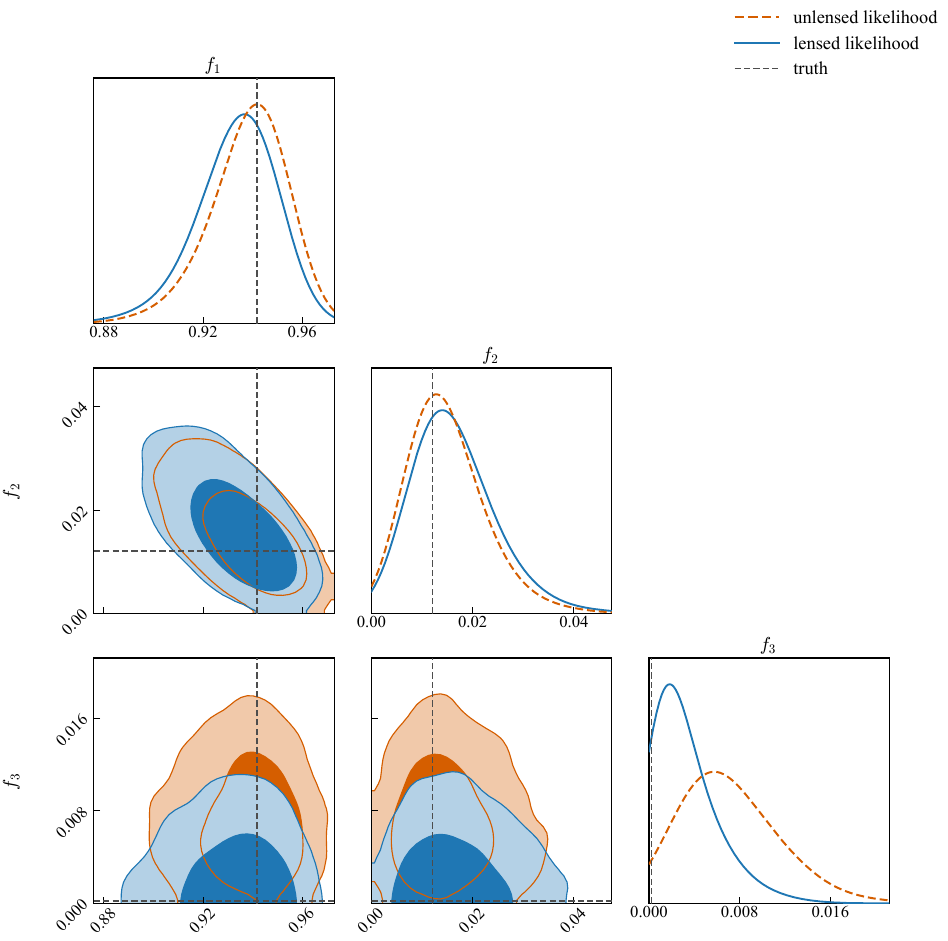}
		\includegraphics[width=0.49\linewidth]{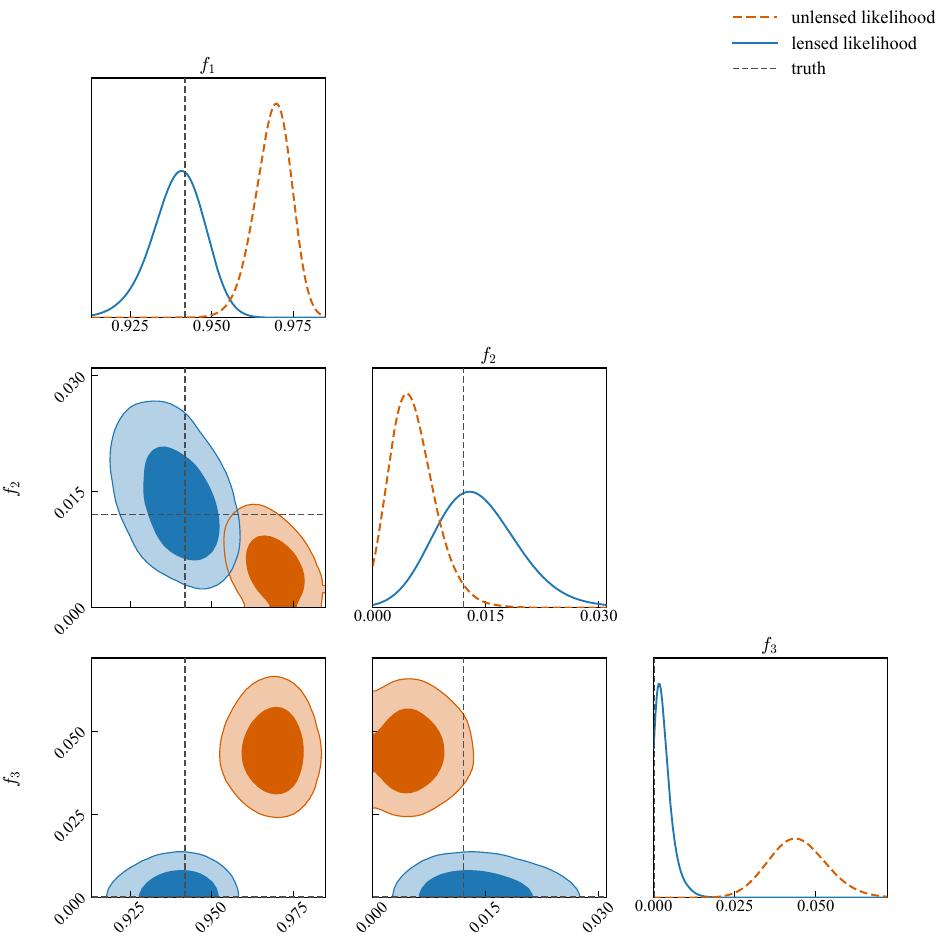}
		\caption{\label{fig:lisa-cond}
			Reference-mixture formation-fraction inference with the conditional likelihood. The left and right panels show the Oguri-VDF and HMF-VDF lens prescriptions, respectively. The contours compare posteriors obtained with unlensed and lensed inference models, using the lensed detectable catalogues as the mock observed data.}
	\end{figure}
	
	\begin{figure}[tbp]
		\centering
		\includegraphics[width=0.49\linewidth]{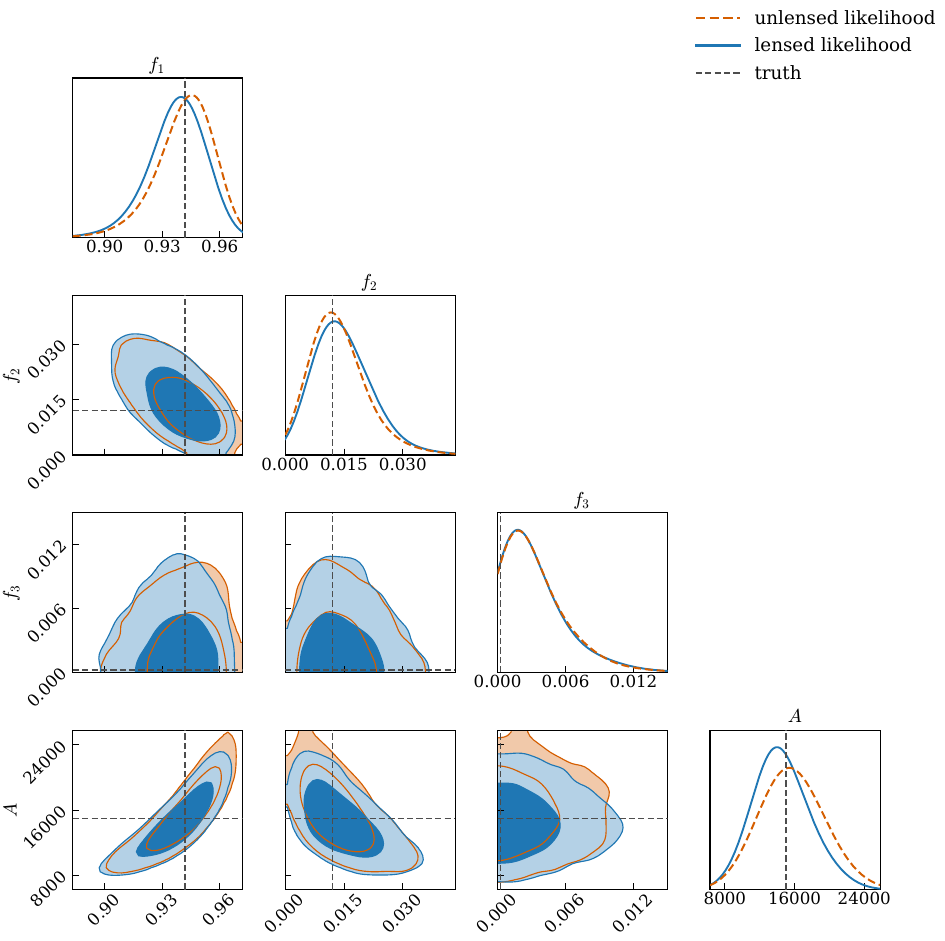}
		\includegraphics[width=0.49\linewidth]{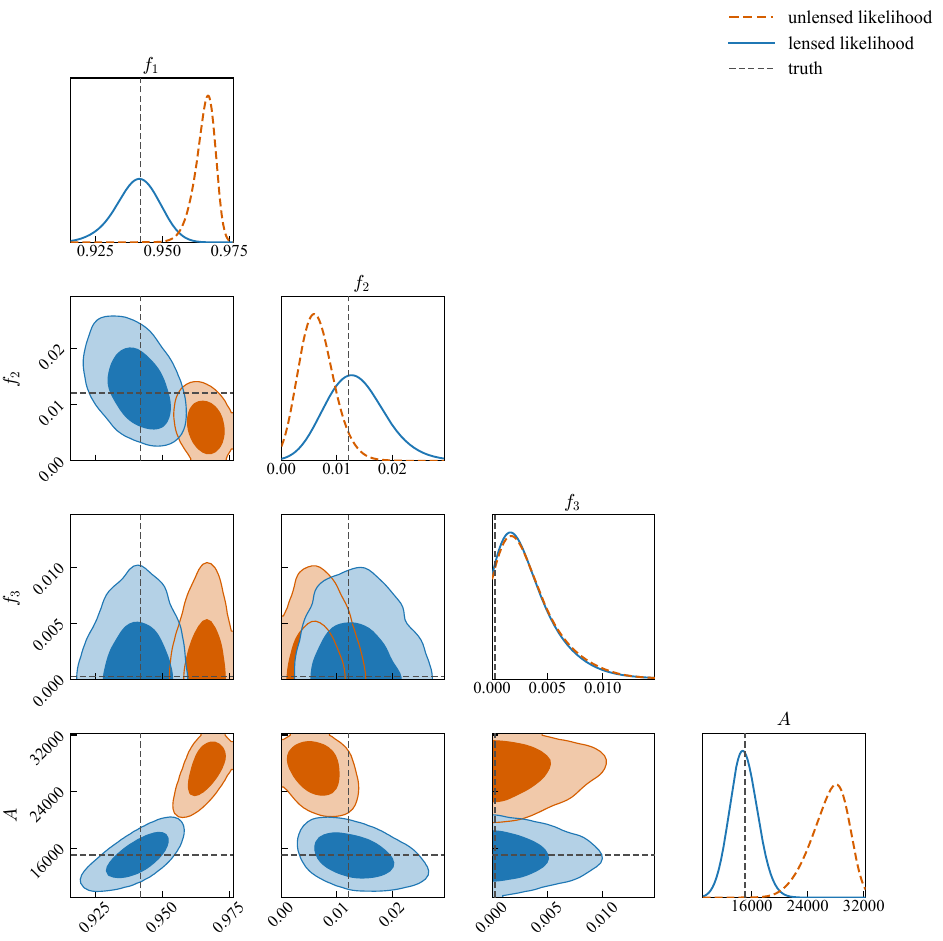}
		\caption{\label{fig:lisa-sampleA}
			Reference-mixture formation-fraction inference with the sampled-$A$ Poisson likelihood. The left and right panels show the Oguri-VDF and HMF-VDF lens prescriptions, respectively. The total intrinsic merger number $A$ is inferred together with the formation fractions; the contours compare unlensed and lensed inference models applied to the lensed detectable catalogues.}
	\end{figure}
	
	Figure~\ref{fig:lisa-cond} shows the reference-mixture results for the conditional likelihood. With the Oguri VDF, the lensing contribution is small, and the unlensed and lensed inference models give similar posterior constraints. The additional lensed detections therefore do not noticeably improve the precision. With the HMF-derived effective VDF, the larger optical depth promotes more subthreshold events into the detectable sample, especially from the light-seed channels. Interpreting these lensed mock data with an unlensed model shifts the recovered fractions, whereas the corresponding lensed model removes most of the mismatch and recovers $f_1$ and $f_2$ within the quoted credible intervals.
	
	Figure~\ref{fig:lisa-sampleA} shows how this comparison changes when the total event count is included but $A$ remains free. The sampled-$A$ likelihood reduces some of the model mismatch, although variations in $A$ can absorb part of the lensing-induced change and limit the gain over the conditional likelihood. The complete numerical comparison in Table~\ref{tab:fiducial-formation-summary} also includes the known-$A$ case. Fixing the intrinsic normalization gives the tightest constraints. When the lensed inference model is used, the recovered $f_1$ and $f_2$ are nearly identical for the Oguri and HMF prescriptions, suggesting that reliable event-number normalization can reduce the sensitivity of these estimates to the lens-population prescription.
	
	\begin{table}[tbp]
		\renewcommand{\arraystretch}{1.25}
		\small
		\setlength{\tabcolsep}{4pt}
	\centering
		\caption{\label{tab:fiducial-formation-summary}
			Reference-mixture LISA posterior summary for the intrinsic formation-fraction parameters. Entries give posterior medians with 68\% credible intervals. The lensed detectable catalogues are used as the mock observed data. The model column indicates whether the inference uses unlensed or lensed inference models.}
		
			\begin{tabular*}{\textwidth}{@{\extracolsep{\fill}}llllll@{}}
				Lens prescription & Likelihood & Model & $f_1$ & $f_2$ & $f_3$ \\
				\hline
				True value & -- & -- & 0.94184 & 0.01205 & 0.00024 \\
				\hline
				Oguri & conditional & unlen & $0.9400^{+0.0110}_{-0.0153}$ & $0.0138^{+0.0074}_{-0.0055}$ & $0.0064^{+0.0046}_{-0.0032}$ \\
				Oguri & conditional & len   & $0.9351^{+0.0119}_{-0.0153}$ & $0.0151^{+0.0082}_{-0.0058}$ & $0.0023^{+0.0032}_{-0.0017}$ \\
				Oguri & sampled-$A$ & unlen & $0.9435^{+0.0107}_{-0.0140}$ & $0.0126^{+0.0066}_{-0.0049}$ & $0.0023^{+0.0032}_{-0.0016}$ \\
				Oguri & sampled-$A$ & len   & $0.9387^{+0.0114}_{-0.0130}$ & $0.0137^{+0.0074}_{-0.0054}$ & $0.0022^{+0.0031}_{-0.0016}$ \\
				Oguri & known-$A$   & unlen & $0.9415^{+0.0019}_{-0.0020}$ & $0.0130^{+0.0047}_{-0.0039}$ & $0.0022^{+0.0031}_{-0.0016}$ \\
				Oguri & known-$A$   & len   & $0.9417^{+0.0020}_{-0.0019}$ & $0.0126^{+0.0044}_{-0.0038}$ & $0.0022^{+0.0031}_{-0.0016}$ \\
				\hline
				HMF   & conditional & unlen & $0.9691^{+0.0043}_{-0.0054}$ & $0.0049^{+0.0029}_{-0.0022}$ & $0.0441^{+0.0079}_{-0.0072}$ \\
				HMF   & conditional & len   & $0.9401^{+0.0063}_{-0.0073}$ & $0.0134^{+0.0052}_{-0.0042}$ & $0.0022^{+0.0028}_{-0.0015}$ \\
				HMF   & sampled-$A$ & unlen & $0.9664^{+0.0024}_{-0.0039}$ & $0.0062^{+0.0028}_{-0.0025}$ & $0.0022^{+0.0031}_{-0.0016}$ \\
				HMF   & sampled-$A$ & len   & $0.9408^{+0.0066}_{-0.0073}$ & $0.0131^{+0.0047}_{-0.0043}$ & $0.0021^{+0.0029}_{-0.0015}$ \\
				HMF   & known-$A$   & unlen & $0.9390^{+0.0019}_{-0.0021}$ & $0.0164^{+0.0053}_{-0.0046}$ & $0.0021^{+0.0031}_{-0.0015}$ \\
				HMF   & known-$A$   & len   & $0.9418^{+0.0018}_{-0.0019}$ & $0.0126^{+0.0043}_{-0.0035}$ & $0.0021^{+0.0030}_{-0.0016}$ \\
			\end{tabular*}
		
	\end{table}
	
	For the weighted mixtures, we focus on the known-$A$ likelihood in Tables~\ref{tab:lsd-formation-summary} and~\ref{tab:hsnod-formation-summary} of Appendix~\ref{app:inference-results}. These cases show that the benefit of lensing-induced additional detections is not universal: it depends on which channel receives the extra events and whether that channel is distinguishable in the inference space. The HS-d contribution remains negligible even after lensing, so the heavy-seed sector is still effectively HS-nod dominated in the present population models. We therefore focus on $f_1$ and $f_2$ when assessing how lensing changes the recoverable formation-fraction information.
	
	These results motivate the next step: instead of fixing the lensing prescription, we ask whether the strength of the lensing contribution can be inferred jointly with the formation fractions.
	
	\subsection{Lensing-rate inference}
	\label{subsec:lensing-rate-inference}
	
	The previous subsection assumed a fixed lens-population prescription. We now ask whether the strength of the strong-lensing contribution can be inferred together with the formation fractions. We introduce the lensing-rate parameter
	\begin{equation}
		R =
		\frac{\Lambda_{\rm lensed\ source}}
		{\Lambda_{\rm intrinsic}},
		\label{eq:lensing-rate-r}
	\end{equation}
	where $\Lambda_{\rm lensed\ source}$ is the expected number of strongly lensed sources and $\Lambda_{\rm intrinsic}$ is the total number of intrinsic mergers over $10\leq z\leq20$ and a 4-year observing time. The purpose of this extension is to test whether the data can distinguish a change in the intrinsic formation fractions from a change in the lensing contribution.
	
	\begin{table}[tbp]
		\renewcommand{\arraystretch}{1.25}
		\small
		\setlength{\tabcolsep}{4pt}
	\centering
		\caption{\label{tab:fiducial-fr-summary}
			Reference-mixture LISA posterior summary for the joint inference of formation fractions and lensing rate. Entries give posterior medians with 68\% credible intervals.}
		
			\begin{tabular*}{\textwidth}{@{\extracolsep{\fill}}llllll@{}}
				Lens prescription & Likelihood & $f_1$ & $f_2$ & $f_3$ & $R$ \\
				\hline
				Oguri true value & -- & 0.94184 & 0.01205 & 0.00024 & 0.00395 \\
				Oguri & conditional & $0.92523^{+0.01679}_{-0.02645}$ & $0.01847^{+0.01166}_{-0.00742}$ & $0.00233^{+0.00325}_{-0.00170}$ & $0.00718^{+0.00706}_{-0.00363}$ \\
				Oguri & known-$A$ & $0.94190^{+0.00187}_{-0.00199}$ & $0.01262^{+0.00443}_{-0.00383}$ & $0.00231^{+0.00313}_{-0.00168}$ & $0.00488^{+0.00294}_{-0.00218}$ \\
				\hline
				HMF true value & -- & 0.94184 & 0.01205 & 0.00024 & 0.05124 \\
				HMF & conditional & $0.92904^{+0.01627}_{-0.01657}$ & $0.01617^{+0.00790}_{-0.00621}$ & $0.00229^{+0.00306}_{-0.00170}$ & $0.06705^{+0.02029}_{-0.02087}$ \\
				HMF & known-$A$ & $0.94181^{+0.00192}_{-0.00184}$ & $0.01259^{+0.00422}_{-0.00365}$ & $0.00227^{+0.00300}_{-0.00166}$ & $0.05186^{+0.00809}_{-0.00744}$ \\
			\end{tabular*}
		
	\end{table}
	
	The reference-mixture results in Table~\ref{tab:fiducial-fr-summary} show that adding $R$ has only a mild impact on the dominant formation fractions. With the conditional likelihood, part of the lensing contribution can be traded against changes in the intrinsic channel mixture, leading to broader or shifted posteriors. This behavior is largely removed in the known-$A$ case, where the total intrinsic event number is fixed. In that limit, $f_1$ and $f_2$ remain close to their input values, and the inferred $R$ is also closer to the true value.
	
	The same reference-mixture results illustrate two reasons why $R$ can be weakly constrained. First, when the true lensing rate is low, changing $R$ produces only a small change in the detectable population. This is seen in the Oguri prescription, where the posterior on $R$ is broad and often shifted above the input value. By contrast, the HMF prescription leaves a larger imprint on the event number and detected channel mixture, and $R$ is recovered more accurately. Second, $R$ is partially degenerate with the intrinsic channel fractions and the event-number normalization. Comparing the conditional and known-$A$ likelihoods shows that fixing $A$ reduces the freedom to absorb lensing-induced changes into the overall normalization or into the channel fractions, thereby improving the recovery of both $R$ and the dominant formation fractions.
	
	The weighted-mixture cases in Tables~\ref{tab:lsd-fr-summary} and~\ref{tab:hsnod-fr-summary} provide a stress test of this behavior rather than independent precision forecasts for $R$. Changing the seed-channel weights also changes the intrinsic event numbers and the detected channel composition, so the lensing contribution may or may not leave a distinguishable imprint beyond the change in the underlying mixture. These cases show a third reason why $R$ can be weakly constrained: additional lensed detections do not automatically provide independent information on the lensing rate. If the extra lensed events populate a channel or parameter-space region that is distinguishable from the detected sample without lensing, they can help identify the lensing contribution. If they mainly overlap with an already dominant component, their effect resembles a small rescaling of that component and the posterior on $R$ remains broad. This explains why the HMF weighted mixtures generally recover $R$ better than the Oguri mixtures, while some high-lensing cases still have broad posteriors.
	
	Overall, the joint analysis shows that the lensing contribution can be constrained only when it leaves a sufficiently strong and distinguishable imprint on the detectable population. Reliable event-count information helps, but it cannot fully compensate for a weak or degenerate lensing signal.
	
	\section{Discussion and conclusions}
	\label{sec:discussion-conclusions}
	
	We have studied how galaxy-scale strong lensing affects the inference of high-redshift MBHB formation-channel fractions. Using semianalytic source catalogues at $10\leq z\leq20$, we considered four channels corresponding to light- and heavy-seed models with delayed and non-delayed mergers. LISA is used as the reference detector for the main analysis; because Taiji has a similar mHz sensitivity band, the LISA results are expected to be qualitatively representative of Taiji for the MBHB populations considered here.
	
	The inference without lensing shows a clear hierarchy in parameter recoverability. The total light-seed fraction $f_1$ is generally well constrained, because light- and heavy-seed systems occupy different regions of the detectable parameter space. The light-seed delay fraction $f_2$ is less precise and depends on the contribution of LS-d events to the detectable sample. The heavy-seed delay fraction $f_3$ is poorly constrained in all cases, because the HS-d channel contributes very few events; within the present population models, the heavy-seed sector is therefore effectively dominated by HS-nod.
	
	Strong lensing changes the inference only when it leaves a sufficiently large and informative imprint on the detectable population. For the conservative Oguri VDF prescription, the lensing contribution is weak and the lensed and unlensed inference models give similar results. For the HMF-derived prescription, lensing promotes more subthreshold events into the detectable sample and can bias the recovered formation fractions if it is ignored. The improvement in statistical precision is not automatic: additional lensed detections help only when they add channel-discriminating information rather than merely increasing the event number of an already dominant component.
	
	Event-count information is central to separating these effects. The conditional likelihood uses only the shape of the detected-event distribution, whereas the Poisson likelihoods also use the total event number. The known-$A$ case is an idealized limit, but it shows that strong external or hierarchical constraints on the intrinsic event-number normalization can reduce the sensitivity of the dominant formation-fraction estimates to the lens-population prescription, provided that lensing is included in the inference model. The same conclusion appears in the joint inference of the lensing-rate parameter $R$: the HMF case gives a clearer and more recoverable lensing signal, while the low-rate Oguri case leaves broad and sometimes biased constraints on $R$.
	
	This work is a controlled catalogue-based population forecast. We have assumed that the relevant formation channels, their detected-event distributions, and their detection efficiencies are known, and we have fixed the individual lens mass model to SIE+shear while bracketing the lens population with two VDF prescriptions. In this work each image is treated as an independent detected event. A source-level treatment may reduce the quantitative impact of lensing but is unlikely to alter the qualitative conclusion that lensing can change the detected channel mixture. In real observations, the channel set, population distributions, event-number normalization, and lens population would need to be inferred or compared within a broader hierarchical framework. Within these assumptions, our results show that galaxy-scale strong lensing and event-count information should be incorporated when using high-redshift MBHB detections to infer intrinsic formation-channel fractions.

\acknowledgments

\begin{sloppypar}
This work was supported by the National Science and Technology Major Project of China
(No. 2024ZD1100601), the National Key R\&D Program of China
(Grant No. 2021YFC2203002), and the National Natural Science Foundation of China
(Grant Nos. 12473075 and 12173071).
\end{sloppypar}

     \clearpage
     \appendix
     \section{Additional event-number and inference results}
     \label{app:inference-results}
	
	This appendix collects the complete event-number and numerical posterior summaries for the weighted mixtures used in Sec.~\ref{sec:population-inference}. The mock observed catalogue and inference model are specified in each caption. The labels ``unlen'' and ``len'' indicate whether the inference model uses unlensed or lensed detected-event distributions.
	
	\begin{table}[htbp!]
		\renewcommand{\arraystretch}{1.1}
		\small
	\centering
		\caption{\label{tab:mixed-detectable-composition}
			LISA-detectable event numbers for the reference and weighted mixtures. Entries give 4-year weighted expected counts for the unlensed, Oguri-VDF, and HMF-VDF cases.}
		
			\begin{tabular*}{\textwidth}{@{\extracolsep{\fill}}lllllll@{}}
				Mixture & Case & Total & LS-d & LS-nod & HS-d & HS-nod \\
				\hline
				\multicolumn{7}{c}{Reference mixture} \\
				\hline
				Reference & unlen & 910.67 & 12.79 & 24.25 & 0.21 & 873.42 \\
				& Oguri & 918.25 & 12.97 & 28.17 & 0.21 & 876.90 \\
				& HMF   & 1007.45 & 15.04 & 74.81 & 0.22 & 917.39 \\
				\hline
				\multicolumn{7}{c}{LS-d-weighted mixtures} \\
				\hline
				Mild & unlen & 315.63 & 39.37 & 7.46 & 0.06 & 268.74 \\
				& Oguri & 318.44 & 39.91 & 8.67 & 0.06 & 269.81 \\
				& HMF   & 351.62 & 46.27 & 23.02 & 0.07 & 282.27 \\
				\cline{1-7}
				Intermediate & unlen & 84.56 & 49.69 & 0.94 & 0.008 & 33.92 \\
				& Oguri & 85.53 & 50.37 & 1.09 & 0.008 & 34.06 \\
				& HMF   & 96.94 & 58.40 & 2.91 & 0.008 & 35.63 \\
				\cline{1-7}
				Extreme & unlen & 55.54 & 50.98 & 0.12 & 0.001 & 4.44 \\
				& Oguri & 56.28 & 51.68 & 0.14 & 0.001 & 4.45 \\
				& HMF   & 64.96 & 59.92 & 0.38 & 0.001 & 4.66 \\
				\hline
				\multicolumn{7}{c}{HS-nod-weighted mixtures} \\
				\hline
				Mild & unlen & 2698.93 & 3.94 & 7.46 & 0.06 & 2687.47 \\
				& Oguri & 2710.91 & 3.99 & 8.67 & 0.06 & 2698.19 \\
				& HMF   & 2850.47 & 4.63 & 23.02 & 0.07 & 2822.76 \\
				\cline{1-7}
				Intermediate & unlen & 3393.36 & 0.50 & 0.94 & 0.008 & 3391.91 \\
				& Oguri & 3407.04 & 0.50 & 1.09 & 0.008 & 3405.44 \\
				& HMF   & 3566.16 & 0.58 & 2.91 & 0.008 & 3562.66 \\
				\cline{1-7}
				Extreme & unlen & 3424.25 & 0.34 & 0.65 & 0.006 & 3423.25 \\
				& Oguri & 3438.01 & 0.35 & 0.76 & 0.006 & 3436.90 \\
				& HMF   & 3598.00 & 0.40 & 2.01 & 0.006 & 3595.58 \\
			\end{tabular*}
		
	\end{table}
	
	\begin{table}[htbp!]
		\renewcommand{\arraystretch}{1.5}
		\scriptsize
		\setlength{\tabcolsep}{3pt}
	\centering
		\caption{\label{tab:baseline-weighted}
			Lensing-free LISA inference for the weighted seed-channel mixtures with the known-$A$ Poisson likelihood. The unlensed detectable catalogues are used as the mock observed data and the inference uses the unlensed model. Entries give posterior medians with 68\% credible intervals.}
		
			\begin{tabular*}{\textwidth}{@{\extracolsep{\fill}}llcccccc@{}}
				Family & Case
				& $f_{1,\rm true}$ & $f_1$
				& $f_{2,\rm true}$ & $f_2$
				& $f_{3,\rm true}$ & $f_3$ \\
				\hline
				LS-d weighted & mild
				& 0.94723 & $0.94716^{+0.00309}_{-0.00320}$
				& 0.10871 & $0.11002^{+0.01957}_{-0.01778}$
				& 0.00024 & $0.00541^{+0.00840}_{-0.00396}$ \\
				& intermediate
				& 0.97261 & $0.97177^{+0.00452}_{-0.00497}$
				& 0.54944 & $0.55844^{+0.08553}_{-0.07482}$
				& 0.00024 & $0.03257^{+0.04957}_{-0.02387}$ \\
				& extreme
				& 0.99412 & $0.99312^{+0.00261}_{-0.00342}$
				& 0.90536 & $0.87495^{+0.08341}_{-0.09995}$
				& 0.00024 & $0.14539^{+0.18697}_{-0.10400}$ \\
				\hline
				HS-nod weighted & mild
				& 0.61827 & $0.61793^{+0.00712}_{-0.00752}$
				& 0.01205 & $0.01423^{+0.00821}_{-0.00678}$
				& $2.39\times10^{-5}$ & $0.00081^{+0.00119}_{-0.00060}$ \\
				& intermediate
				& 0.13941 & $0.13772^{+0.01428}_{-0.01468}$
				& 0.01205 & $0.03248^{+0.03843}_{-0.02236}$
				& $2.39\times10^{-6}$ & $0.00066^{+0.00098}_{-0.00048}$ \\
				& extreme
				& 0.09998 & $0.09688^{+0.01455}_{-0.01591}$
				& 0.01205 & $0.04166^{+0.04962}_{-0.02835}$
				& $1.64\times10^{-6}$ & $0.00061^{+0.00091}_{-0.00044}$ \\
			\end{tabular*}
		
	\end{table}

	\begin{table}[tbp]
		\renewcommand{\arraystretch}{1.5}
		\small
		\setlength{\tabcolsep}{4pt}
	\centering
		\caption{\label{tab:lsd-formation-summary}
			Formation-fraction inference for the LS-d-weighted mixtures using the known-$A$ Poisson likelihood with the lensed inference model. Entries give posterior medians with 68\% credible intervals.}
		
			\begin{tabular*}{\textwidth}{@{\extracolsep{\fill}}llllll@{}}
				Mixture & Lens & $f_{1,\rm true}$ & $f_1$ & $f_{2,\rm true}$ & $f_2$ \\
				\hline
				Mild & Oguri & 0.94723 & $0.9471^{+0.0031}_{-0.0033}$ & 0.10871 & $0.1102^{+0.0187}_{-0.0178}$ \\
				Mild & HMF   & 0.94723 & $0.9471^{+0.0030}_{-0.0032}$ & 0.10871 & $0.1101^{+0.0174}_{-0.0152}$ \\
				Intermediate & Oguri & 0.97261 & $0.9720^{+0.0044}_{-0.0052}$ & 0.54944 & $0.5571^{+0.0816}_{-0.0768}$ \\
				Intermediate & HMF   & 0.97261 & $0.9720^{+0.0043}_{-0.0048}$ & 0.54944 & $0.5539^{+0.0773}_{-0.0724}$ \\
				Extreme & Oguri & 0.99412 & $0.9932^{+0.0026}_{-0.0034}$ & 0.90536 & $0.8761^{+0.0776}_{-0.0963}$ \\
				Extreme & HMF   & 0.99412 & $0.9934^{+0.0025}_{-0.0033}$ & 0.90536 & $0.8822^{+0.0711}_{-0.0929}$ \\
			\end{tabular*}
		
	\end{table}
	
	\begin{table}[tbp]
		\renewcommand{\arraystretch}{1.5}
		\small
		\setlength{\tabcolsep}{4pt}
	\centering
		\caption{\label{tab:hsnod-formation-summary}
			Formation-fraction inference for the HS-nod-weighted mixtures using the known-$A$ Poisson likelihood with the lensed inference model. Entries give posterior medians with 68\% credible intervals.}
		
			\begin{tabular*}{\textwidth}{@{\extracolsep{\fill}}llllll@{}}
				Mixture & Lens & $f_{1,\rm true}$ & $f_1$ & $f_{2,\rm true}$ & $f_2$ \\
				\hline
				Mild & Oguri & 0.61827 & $0.6179^{+0.0076}_{-0.0076}$ & 0.01205 & $0.0144^{+0.0088}_{-0.0067}$ \\
				Mild & HMF   & 0.61827 & $0.6180^{+0.0069}_{-0.0073}$ & 0.01205 & $0.0142^{+0.0076}_{-0.0065}$ \\
				Intermediate & Oguri & 0.13941 & $0.1376^{+0.0149}_{-0.0147}$ & 0.01204 & $0.0311^{+0.0392}_{-0.0208}$ \\
				Intermediate & HMF   & 0.13941 & $0.1379^{+0.0147}_{-0.0145}$ & 0.01204 & $0.0296^{+0.0354}_{-0.0196}$ \\
				Extreme & Oguri & 0.09998 & $0.0967^{+0.0160}_{-0.0159}$ & 0.01205 & $0.0408^{+0.0541}_{-0.0280}$ \\
				Extreme & HMF   & 0.09998 & $0.0978^{+0.0149}_{-0.0150}$ & 0.01205 & $0.0369^{+0.0483}_{-0.0251}$ \\
			\end{tabular*}
		
	\end{table}
	
	\begin{table}[tbp]
		\renewcommand{\arraystretch}{1.5}
		\scriptsize
		\setlength{\tabcolsep}{3pt}
	\centering
		\caption{\label{tab:lsd-fr-summary}
			Joint inference of formation fractions and lensing rate for the LS-d-weighted mixtures using the known-$A$ Poisson likelihood with the lensed inference model. Entries give posterior medians with 68\% credible intervals.}
		
			\begin{tabular*}{\textwidth}{@{\extracolsep{\fill}}llllllll@{}}
				Mixture & Lens & $R_{\rm true}$ & $R$ & $f_{1,\rm true}$ & $f_1$ & $f_{2,\rm true}$ & $f_2$ \\
				\hline
				Mild & Oguri & 0.00394 & $0.00677^{+0.00645}_{-0.00404}$ & 0.94723 & $0.9473^{+0.0034}_{-0.0032}$ & 0.10871 & $0.1098^{+0.0183}_{-0.0171}$ \\
				Mild & HMF   & 0.05094 & $0.05250^{+0.01365}_{-0.01252}$ & 0.94723 & $0.9472^{+0.0031}_{-0.0035}$ & 0.10871 & $0.1094^{+0.0183}_{-0.0160}$ \\
				Intermediate & Oguri & 0.00385 & $0.01415^{+0.01858}_{-0.00987}$ & 0.97261 & $0.9724^{+0.0044}_{-0.0052}$ & 0.54944 & $0.5393^{+0.0809}_{-0.0740}$ \\
				Intermediate & HMF   & 0.04950 & $0.05232^{+0.02491}_{-0.02157}$ & 0.97261 & $0.9723^{+0.0044}_{-0.0051}$ & 0.54944 & $0.5510^{+0.0805}_{-0.0725}$ \\
				Extreme & Oguri & 0.00378 & $0.02164^{+0.02584}_{-0.01499}$ & 0.99412 & $0.9934^{+0.0025}_{-0.0034}$ & 0.90536 & $0.8435^{+0.0930}_{-0.1080}$ \\
				Extreme & HMF   & 0.04829 & $0.05581^{+0.02726}_{-0.02539}$ & 0.99412 & $0.9931^{+0.0026}_{-0.0034}$ & 0.90536 & $0.8670^{+0.0830}_{-0.0948}$ \\
			\end{tabular*}
		
	\end{table}
	
	\begin{table}[tbp]
		\renewcommand{\arraystretch}{1.5}
		\scriptsize
		\setlength{\tabcolsep}{3pt}
	\centering
		\caption{\label{tab:hsnod-fr-summary}
			Joint inference of formation fractions and lensing rate for the HS-nod-weighted mixtures using the known-$A$ Poisson likelihood with the lensed inference model. Entries give posterior medians with 68\% credible intervals.}
		
			\begin{tabular*}{\textwidth}{@{\extracolsep{\fill}}llllllll@{}}
				Mixture & Lens & $R_{\rm true}$ & $R$ & $f_{1,\rm true}$ & $f_1$ & $f_{2,\rm true}$ & $f_2$ \\
				\hline
				Mild & Oguri & 0.00385 & $0.00685^{+0.00645}_{-0.00408}$ & 0.61827 & $0.6197^{+0.0077}_{-0.0073}$ & 0.01205 & $0.0139^{+0.0088}_{-0.0068}$ \\
				Mild & HMF   & 0.04949 & $0.05037^{+0.01421}_{-0.01165}$ & 0.61827 & $0.6187^{+0.0085}_{-0.0081}$ & 0.01205 & $0.0137^{+0.0083}_{-0.0063}$ \\
				Intermediate & Oguri & 0.00370 & $0.01379^{+0.01849}_{-0.00955}$ & 0.13941 & $0.1488^{+0.0195}_{-0.0186}$ & 0.01205 & $0.0268^{+0.0339}_{-0.0184}$ \\
				Intermediate & HMF   & 0.04689 & $0.04612^{+0.02388}_{-0.02071}$ & 0.13941 & $0.1385^{+0.0224}_{-0.0230}$ & 0.01205 & $0.0281^{+0.0349}_{-0.0189}$ \\
				Extreme & Oguri & 0.00368 & $0.01527^{+0.02031}_{-0.01063}$ & 0.09998 & $0.1101^{+0.0216}_{-0.0182}$ & 0.01205 & $0.0335^{+0.0485}_{-0.0233}$ \\
				Extreme & HMF   & 0.04668 & $0.04267^{+0.02476}_{-0.02076}$ & 0.09998 & $0.0944^{+0.0244}_{-0.0221}$ & 0.01205 & $0.0376^{+0.0508}_{-0.0256}$ \\
			\end{tabular*}
		
	\end{table}
	
	\FloatBarrier
	\clearpage

	% Add references with:
	\bibliographystyle{JHEP}
	\bibliography{references}
	
\end{document}